\documentclass[acmlarge,natbib=false,manuscript,screen,nonacm]{acmart}
\setcopyright{none}
\acmDOI{}
\acmISBN{}
\acmConference[]{}{}{}
\renewcommand{\footnotetextcopyrightpermission}[1]{}

\AtBeginDocument{%
  \providecommand\BibTeX{{%
    \normalfont B\kern-0.5em{\scshape i\kern-0.25em b}\kern-0.8em\TeX}}}

\RequirePackage[
  datamodel=acmdatamodel,
  style=acmnumeric,
  ]{biblatex}

\usepackage[nameinlink]{cleveref} 

\usepackage[nolist,nohyperlinks]{acronym} 
\begin{acronym}
    \acro{SIEM}{Security Information and Event Management}
    \acro{IDS}{Intrusion Detection System}
    \acro{IPS}{Intrusion Prevention System}
    \acro{IDPS}{Intrusion Detection and Prevention System}
    \acro{APT}{Advanced Persistent Threat}
    \acro{ML}{Machine Learning}
    \acro{DL}{Deep Learning}
    \acro{EDR}{Endpoint Detection and Response}
    \acro{SOAR}{Security Orchestration Automation and Responses}
    \acro{EPP}{Endpoint Security and Protection Platforms}
    \acro{IAM}{Identity and Access Management}
    \acro{C2}{Command-and-Control}
    \acro{HDBSCAN}{Hierarchical Density-Based Spatial Clustering of Applications with Noise}
    \acro{LSTM}{Long Short-term Memory}
    \acro{CNN}{Convolutional Neural Network}
\end{acronym}

\newcommand{\citet}[1]{\citeauthor{#1}~\cite{#1}}

\begin{document}

\title[Exploring AI-Enabled Cybersecurity Frameworks]{Exploring AI-Enabled Cybersecurity Frameworks: Deep-Learning Techniques, GPU Support, and Future Enhancements}

\author{Tobias Becher}
\email{t.becher.1@campus.tu-berlin.de}
\orcid{0009-0008-9482-3329}
\author{Simon Torka}
\email{simon.torka@dai-labor.de}
\orcid{0009-0009-1201-1915}
\affiliation{%
  \institution{DAI-Lab, Technische Universität Berlin}
  \streetaddress{Ernst-Reuter-Platz 7}
  \city{Berlin}
  \country{Germany}
  \postcode{10587}
}



\begin{abstract}
    Traditional rule-based cybersecurity systems have proven highly effective against known malware threats. However, they face challenges in detecting novel threats. To address this issue, emerging cybersecurity systems are incorporating AI techniques, specifically deep-learning algorithms, to enhance their ability to detect incidents, analyze alerts, and respond to events. While these techniques offer a promising approach to combating dynamic security threats, they often require significant computational resources. Therefore, frameworks that incorporate AI-based cybersecurity mechanisms need to support the use of GPUs to ensure optimal performance.

    Many cybersecurity framework vendors do not provide sufficiently detailed information about their implementation, making it difficult to assess the techniques employed and their effectiveness. This study aims to overcome this limitation by providing an overview of the most used cybersecurity frameworks that utilize AI techniques, specifically focusing on frameworks that provide comprehensive information about their implementation. Our primary objective is to identify the deep-learning techniques employed by these frameworks and evaluate their support for GPU acceleration. We have identified a total of \emph{two} deep-learning algorithms that are utilized by \emph{three} out of 38 selected cybersecurity frameworks. Our findings aim to assist in selecting open-source cybersecurity frameworks for future research and assessing any discrepancies between deep-learning techniques used in theory and practice.
\end{abstract}



\maketitle

\section{Introduction}
\label{sec:intro}

As cybersecurity threats become more prevalent~\cite{bundeskriminalamt_filed_2021}, cybersecurity tools must retain the ability to keep systems safe. Traditionally, cyber threats have been combated using a rules-based approach that involves investigating detected incidents, classifying attack vectors, and defining a custom solution to mitigate the threat. For example, part of the initial strategy to stop one of the first computer worms, the \textit{Morris Worm}, was to put simple rules in place. In this case, disabling the \textit{FINGER-Daemon} that the worm relied on eliminated one of the three ways it could infect a system~\cite{jajoo_study_2021}. Over time, threats have become increasingly complex, and of course, security solutions must adapt~\cite{berman_survey_2019}. 

Cybersecurity vendors now offer a myriad of different tools and platforms to handle security, privacy, and safety concerns. By examining the market's key players, as we describe in \Cref{ssec:methodology}, it is possible to uncover any potential gap between theory and practice, especially in their use of novel techniques. However, assessing the technical details of cybersecurity products can be a challenging task. Publicly available documentation is often insufficient, making it difficult to conduct a thorough evaluation. In the materials we reviewed to evaluate cybersecurity products from the major vendors in the marketplace, we encountered several instances where a lack of detail made it difficult to determine the underlying technical specifications of a product. For example, the product materials for Fortinet\footnote{\url{https://www.fortinet.com/}} and SentinelOne\footnote{\url{https://www.sentinelone.com/}}, two multinational cybersecurity companies that are significant vendors in the market, do not provide enough detail for a thorough review.
Fortinet offers products categorized as network, application, and endpoint security, among others. However, it is not immediately clear how these solutions operate internally, as Fortinet does not provide implementation details to the public~\cite{fortinet_fortigate_2023}. The publicly available resources for the AI-powered security tool \emph{Singularity}, offered by SentinelOne, include references to a mix of a rules-based and an AI-powered approach~\cite{sentinelone_sentinelone_nodate}. Since most cybersecurity frameworks are closed-source, private software solutions, it is difficult to gain more detailed insight into their inner workings. 

Despite often not providing technical detail, available product information makes it evident that marketing campaigns are now emphasizing the AI features of software solutions~\cite{sentinelone_sentinelone_nodate,vectra_ai_2022,ossec_get_nodate,allen_fingerprinting_2021,fortinet_fortigate_2023}. This serves as another indication that, although the traditional approach still holds value, recent advances in artificial intelligence have opened new avenues for combating cyber threats. This is particularly true in the areas of intrusion detection, phishing and spam detection, threat intelligence, and user behavior analysis~\cite{kinyua_aiml_2021}. Intrusion detection, for example, can use machine-learning algorithms to distinguish between normal and abnormal traffic on a network. With a well-trained machine learning model and high-quality network data, modern cybersecurity systems should be able to detect the spread of a novel virus without having seen similar malware before. A Morris Worm redux, i.e., an entirely novel malware threat, should be detected and acted upon preemptively. Additionally, the use of AI could also improve a cybersecurity product's ability to detect modern, complex multi-step attacks through semantic analysis~\cite{luh_semantics-aware_2017}.

Of course, this solution is not without challenges or caveats. Typically, these systems include the potential for a high rate of false alarms~\cite{xin_machine_2018} and can be very computationally intensive, especially for deep-learning methods~\cite{shrestha_review_2019}. While this approach may be deemed appropriate for security-critical infrastructure that has ample resources, it is a concern for a variety of other use cases. This is especially true for private individuals and small and medium-sized enterprises, which often lack the infrastructure to provide sufficient computing resources for larger deep-learning models. To make the most efficient use of hardware, cybersecurity solutions that rely on deep learning must use GPUs for training and inference~\cite{shrestha_review_2019}.

Naturally, this raises the question of what solutions currently exist, what deep-learning capabilities they have, and whether they make efficient use of hardware, i.e., whether they enable GPU-supported deep learning. The relevance for both researchers and security vendors is thus clear, as answering these questions can provide a clear direction for future improvements.


This leads us to the central questions for this study. 
\begin{itemize}
\setlength\itemsep{0em}
    \item[RQ1.] What information is available about the techniques employed by the most used cybersecurity frameworks? 
    \item[RQ2.] Which frameworks are open-source and could be used in academic research?
    \item[RQ3.] What deep learning techniques are used in existing cybersecurity frameworks? How is their use supported by GPU resources?
\end{itemize}

In exploring answers to these questions, our contribution is twofold. First, we provide an overview of the techniques used in current cybersecurity frameworks. Second, we aid researchers in selecting cybersecurity frameworks for future studies. 
Although we try to provide a wide range of techniques and frameworks, we do not claim to be exhaustive. This is due to two facts. First, the cybersecurity landscape is constantly evolving, and commercial solutions generally do not provide enough information to be thoroughly vetted. Second, there are recent studies that, while not focusing on cybersecurity frameworks, analyze the cybersecurity market and its players, providing a more complete overview of the market. We will discuss these in \Cref{sec:stateoftheart}.

In this paper, we adopt a systematic approach to exploring cybersecurity frameworks, with particular attention to their support for GPU acceleration and possible enhancements. We begin with a thorough examination of publicly available documentation for existing cybersecurity frameworks, including various types of cybersecurity solutions such as \ac{SIEM} systems, \acp{IDS}, and other related software. Our goal is to gain insight into the techniques used by these frameworks in handling various security tasks. We also investigate the extent to which GPUs are supported in training, testing, and inference of deep-learning-based algorithms in these frameworks.
Exploring cybersecurity frameworks with GPU support allows us to understand the state of the art and propose practical enhancements bridging the gap between theory and practice. 
Ultimately, this research holds significant importance as it represents a step toward developing resilient cybersecurity solutions. By leveraging the right deep-learning solutions, we can contribute to the development of robust and resilient cybersecurity solutions that can better protect organizations and individuals from cyberattacks.

This paper is structured as follows. \Cref{sec:stateoftheart} lays out related research in the field and differences to our work are pointed out. Next, in \Cref{sec:background}, we will cover our methodology for selecting and reviewing relevant information, as well as discuss related terminology and the challenges of using deep learning for cybersecurity. Afterward, we will move on to \Cref{sec:frameworks}, where we will delve into the various frameworks we have studied and the techniques they employ. Finally, in \Cref{sec:conclusion} we conclude and provide a brief outlook on future research on this topic.

\section{State of the Art}
\label{sec:stateoftheart}

To discuss the current state of the art, we divide the related research into two parts. The first part, Cybersecurity Systems, addresses current implementations of various cybersecurity tools, such as antivirus suites, as well as proposed implementations from research. In the second part, we review research that provides an overview of current deep-learning techniques in cybersecurity. As far as we are aware, there is no recent work that addresses GPU support for AI-based cybersecurity tools.

Concerning our study, the first part gives us an overview of market solutions but does not explicitly examine the (deep-learning) techniques used among them. Often, the tools are discussed in a market context and less from a technical point of view.
Compared to the papers in Part Two, which presented deep-learning techniques, we focus on the implementation of these techniques in current cybersecurity software and how well these frameworks can execute them effectively and efficiently.

\paragraph{Cybersecurity Systems.}
In 2020, \citeauthor{tselios_comprehensive_2020} conducted a technical survey investigating the cybersecurity market. They categorized existing cybersecurity solutions into 13 categories, such as \textit{Intrusion Detection and Prevention Systems}, \textit{Endpoint Detection and Response}, or \textit{Identity and Access Management}.  
They concluded that the current cybersecurity market does not offer a product that is capable of holistically covering all capabilities to be effective in all security categories, but that every framework must adopt the latest innovations~\cite{tselios_comprehensive_2020}. 

A study by \citeauthor{gonzalez-granadillo_security_2021} focused on the analysis of the \ac{SIEM} market. They compiled a list of current commercial solutions and classified them according to their functionality and market position. Especially noteworthy are their remarks on the use of machine learning in these \acp{SIEM}. They state that, while machine learning approaches should be used more often, only a few current software solutions include them~\cite{gonzalez-granadillo_security_2021}.

\paragraph{Overview of Deep-Learning Techniques in Cybersecurity.}
The work of \citeauthor{berman_survey_2019} gives a thorough overview of deep-learning techniques in cybersecurity research through January 2019. They describe the techniques and collect usage statistics in the surveyed work. They see the limitations of the field primarily in the lack of quality datasets for training and evaluation, and the fact that these techniques are only used in isolation, i.e., they do not consider the entire attack lifecycle~\cite{berman_survey_2019}.

Other studies have listed various deep-learning algorithms for cybersecurity~\cite{sarker_deep_2021,li_cyber_2018,li_deep_2021,dixit_deep_2021,xin_machine_2018}, either adding works that are not included in the survey by \citeauthor{berman_survey_2019} or discussing other aspects, such as adversarial learning~\cite{li_cyber_2018}.

For example, \citeauthor{dixit_deep_2021} added the category of reinforcement learning algorithms and their use in cybersecurity, while also extending the included research to papers published before May 2020~\cite{dixit_deep_2021}. Taken together, there is a wide range of deep learning techniques being applied to cybersecurity problems.


\section{Background}
\label{sec:background}

Before discussing cybersecurity frameworks, we want to clarify the most important terms used in this paper, our methodology, and finally give a short overview of GPU support in AI-based software. 

\subsection{Methodology}
\label{ssec:methodology}

Since this work reviews various cybersecurity frameworks, it is important to discuss the selection method of software solutions.


According to \citet{haleliuk_why_2023}, IT-Harvest Dashboard, the largest cybersecurity vendor database, counts more than 3,200 companies in the cybersecurity market. The market, while containing several key players, is not dominated by one company~\cite{statista_cybersecurity_nodate}. Therefore, we want to provide our methodology for selecting a representative sample for our survey. We reviewed the products of the key players in the market and any vendors listed in the categories \textit{Intrusion Detection and Prevention Systems}, \textit{Endpoint Detection and Response}, and \textit{Identity and Access Management} in the survey by \citet{tselios_comprehensive_2020}. Additionally, we conducted a Google and Google Scholar search using combinations of the term ``cybersecurity framework'' and one of the terms ``solution'', ``platform'', ``IDS'', ``SIEM'', and ``SOAR'' and their unabbreviated versions. We limited our results to the first three pages of the Google search results and the first page of the Google Scholar results. We selected all results that demonstrated the potential to fit our definition of cybersecurity frameworks, as defined in \Cref{ssec:terminology}. Products that did not provide cybersecurity solutions in the aforementioned fields were discarded. Research works that did not provide implementations were discarded as well. Combining the results from our search with the list of vendors from the survey and market statistics, we arrived at a list of 38 vendors and \emph{no} research works. The lack of research is not necessarily surprising, as cybersecurity frameworks, such as those described in a study by \citet{luh_semantics-aware_2017}, do not meet our criteria for cybersecurity frameworks, primarily because of their lack of extensibility.

The selected vendors' websites were searched for product information, such as white papers and technical documentation, and reviewed for machine- and deep-learning features. It is important to be aware of the fact that the analysis presented here is based on information that may not have been specifically crafted for a scientific or technically proficient audience. As a result, there is a possibility that the reviewed materials, including marketing materials, may not accurately convey all of the technical aspects of the product, potentially resulting in different interpretations and decreased reliability. However, given the lack of other research on the topic, we find this to be the best available method for analyzing these products.

\subsection{Terminology}
\label{ssec:terminology}

First, we want to disambiguate the term cybersecurity framework from related vocabulary. Then, we give brief definitions for the three categories of cybersecurity systems. 

In this study, we use the terms \textit{cybersecurity framework} and \textit{cybersecurity platform} interchangeably to describe an extensible software framework that is tasked with achieving one or more of the cybersecurity goals of confidentiality, integrity, and availability. Our definition is not to be confused with the commonly accepted meaning of the term in the context of organizational security. There, cybersecurity frameworks are often also referred to as cybersecurity risk management frameworks, as e.g., examined by \citet{taherdoost_understanding_2022}, and represent a distinct area of study and practice.

\begin{definition}
\label{def:framework}
    A \emph{cybersecurity framework} refers to a software system designed to manage and safeguard digital assets from potential threats and attacks. It serves as a foundational architecture that incorporates various cybersecurity tools. Therefore, it can contain capabilities of more specific cybersecurity tools of categories such as \acl{EDR}, \acl{IDS}, \acl{SIEM}, and \acl{SOAR}, allowing for their seamless integration and collaborative utilization to enhance overall security.
\end{definition}

Hence, our chosen definition of the term \emph{cybersecurity framework} is a flexible one to encompass a variety of different approaches to cybersecurity, if they can be extended by additional techniques. For example, we do not consider a traditional firewall a cybersecurity framework, since it is typically not possible to extend its capabilities without it becoming a different cybersecurity solution, e.g., an \ac{EDR} tool.  
A \emph{cybersecurity solution} can be any kind of cybersecurity-related software solution and therefore includes cybersecurity frameworks as well.

\subsection{GPU Support for Machine-Learning Algorithms}
\label{ssec:GPUsupport}

Typically, deep-learning algorithms require GPUs to compute efficiently, with GPUs achieving up to a 10-times acceleration of training time and/or inference~\cite{baykal_comparing_2018}. Popular deep-learning frameworks, like TensorFlow\footnote{\label{fn:tensorflow}\url{https://www.tensorflow.org/}} and PyTorch\footnote{\label{fn:pytorch}\url{https://pytorch.org/}} provide native GPU support. 

While popular \ac{ML} frameworks, such as \emph{Scikit-learn}\footnote{\url{https://scikit-learn.org/stable/}}, do not always provide GPU support, this can often be mitigated by third-party libraries. A prominent example is the RAPIDS\footnote{\url{https://rapids.ai/ecosystem/}} ecosystem by NVIDIA, which provides libraries for executing data science code on GPUs while keeping the same API. For example, to use an \ac{ML} algorithm included in the Scikit-learn library with GPU acceleration, \textit{cuML}\footnote{\url{https://github.com/rapidsai/cuml}} can be used. According to research from 2022, the RAPIDS algorithms are currently not performing as well as their Scikit-learn counterparts when used with default settings but can still provide satisfactory results when the hyperparameters are tuned accordingly~\cite{motylinski_gpu-based_2022}.

\subsection{Challenges of Deep Learning in Cybersecurity}
\label{ssec:dlchallenges}

The challenges for applying deep learning are similar across computer science domains and contain aspects such as the quality of input data, data representation in feature space, training time, inference performance, and more. Focusing on aspects that are relevant to the use of deep-learning algorithms in cybersecurity frameworks, we briefly discuss the distinctions these challenges have in cybersecurity over other disciplines.

\paragraph{Datasets.}
Quality data are a pillar of effective machine learning in every domain. To obtain training data for a cybersecurity machine-learning approach, researchers are presented with a choice: collect the data from real computer systems and/or networks, synthesize a dataset, or use an existing public dataset. Unfortunately, all of these options come with significant challenges, and obtaining enough quality data remains one of the major concerns in cybersecurity research that makes use of machine learning. 

First, collecting real data requires permission from a sufficiently large institution to monitor their activities. This often entails various concerns, such as privacy concerns or concerns over interference with normal operations. This approach is only feasible for projects that require only a very limited scope of data, such as monitoring the network traffic of one device in an isolated network. This way, concerns can be handled on an individual basis. 

Second, synthesizing cybersecurity data is not a trivial task, as it requires extensive knowledge and resources to mimic realistic data. The latest approaches try to synthesize attack data from existing public datasets~\cite{yan_automatically_2019,bourou_review_2021}, and have therefore similar concerns to the third option, using public datasets. Works that aim to emulate blue and red team behavior, i.e., the behavior of defenders and attackers, can contribute to creating these datasets~\cite{applebaum_intelligent_2016,yoo_cyber_2020}. But, as far as we are aware, no complete datasets have been synthesized using these techniques. 

Lastly, using public datasets is the most common choice for researchers~\cite{ferrag_deep_2020}. Even though there are at least 31 published cybersecurity datasets, according to a study by \citeauthor{ferrag_deep_2020}, most research uses only four of them~\cite{ferrag_deep_2020}. This includes the popular KDD Cup 1999 dataset~\cite{kdd_kdd_nodate}, which has been criticized since its inception~\cite{mchugh_testing_2000,siddique_kdd_2019}, as well as the NSL-KDD~\cite{tavallaee_detailed_2009}, UNSW-NB15~\cite{moustafa_unsw-nb15_2015} and CICIDS2017~\cite{sharafaldin_toward_2018} datasets. Criticisms include the recency of the data, as cybersecurity is a fast-evolving field and attacks and defenses are constantly changing, as well as criteria such as data balancing~\cite{sommer_outside_2010,moustafa_evaluation_2016,shaukat_performance_2020}.

\paragraph{Performance.}
While not all cybersecurity tasks must prove temporal adequacy, specific tasks must react in real-time. For example, while the analysis of presumed malware allows for a more lenient time allotment, the reaction to a network attack, like a Denial-Of-Service attack, must be immediate to avoid system failure. If machine-learning algorithms are used to detect such attacks, their inference must fit this demand. The time complexity of the chosen algorithm should therefore be taken into consideration~\cite{shaukat_performance_2020}.

\paragraph{Continual Learning.}
Cybersecurity is highly dynamic, with new threats, known as zero-day exploits, emerging in rapid succession. If an algorithm is not able to continuously learn to detect new threats, it will quickly become obsolete and ineffective. If training consumes a lot of resources, as is typically the case with deep-learning techniques, it is infeasible to train new iterations of the model from scratch. Models must therefore be able to incrementally update their knowledge throughout their lifetime.

\paragraph{Federated Learning.}
To preserve privacy~\cite{mohassel_secureml_2017} and allow rapid access to real-time data for fast model training~\cite{chen_distributed_2021}, the security infrastructure can be distributed onto multiple components. Distribution can also improve and ensure failure safety. However, solving issues such as efficient node connection and communication is a challenging task and central to the success of the distributed machine-learning system~\cite{chen_distributed_2021}.

\paragraph{Result Visualization.}
Results generated by the various algorithms must ultimately inform a human analyst. The level of trust in a cybersecurity system is expected to remain relatively unaltered, even with the implementation of more effective automation, as the maximum extent of trust that can be placed in the system is inherently limited. This is because, security systems can also be targets of attacks, such as adversarial attacks on a neural network tasked with classifying network traces as normal or anomalous. Therefore, analysts have to be able to understand the path from raw network data to intermediate representation in the model's feature space, to the classification as an alert.

Cyber visualizations are being criticized as either too complicated or too basic, as well as too rigid to adapt to different scenarios~\cite{staheli_visualization_2014}. Though the research on visualizations has grown and approaches have been categorized, a standard for cybersecurity visualization has not yet been created~\cite{komadina_analysis_2022}.

\section{Cybersecurity Frameworks}
\label{sec:frameworks}

We reviewed publicly available information on cybersecurity products from 38 different vendors and one research paper and grouped them into six categories, which can be found in \Cref{tab:prodcategories}.

\begin{table}[ht]
    \centering
    \caption{Categories of cybersecurity product information.}
    \label{tab:prodcategories}
    \begin{tabular}{@{}ll@{}}
        \toprule
        \textbf{Category} & \textbf{Description}                                                                                  \\ \midrule
        Rules-based       & The product information hints at using a \emph{rules-based approach} instead of an ML approach.              \\[0.1cm]
        ML                & The product information mentions the use of \emph{machine learning}.                                         \\[0.1cm]
        DL                & The product information mentions the use of \emph{deep learning}.                                            \\[0.1cm]
        AI                & The product information clearly states the use of \emph{AI} but does not elaborate further.                  \\[0.1cm]
        Other             & The product was initially selected but later discarded.                                               \\[0.1cm]
        No info           & The product information does not contain information pertaining to the use of AI.                     \\ \bottomrule
    \end{tabular}
\end{table}

\begin{figure}
    \centering
        \includegraphics[width=.5\columnwidth]{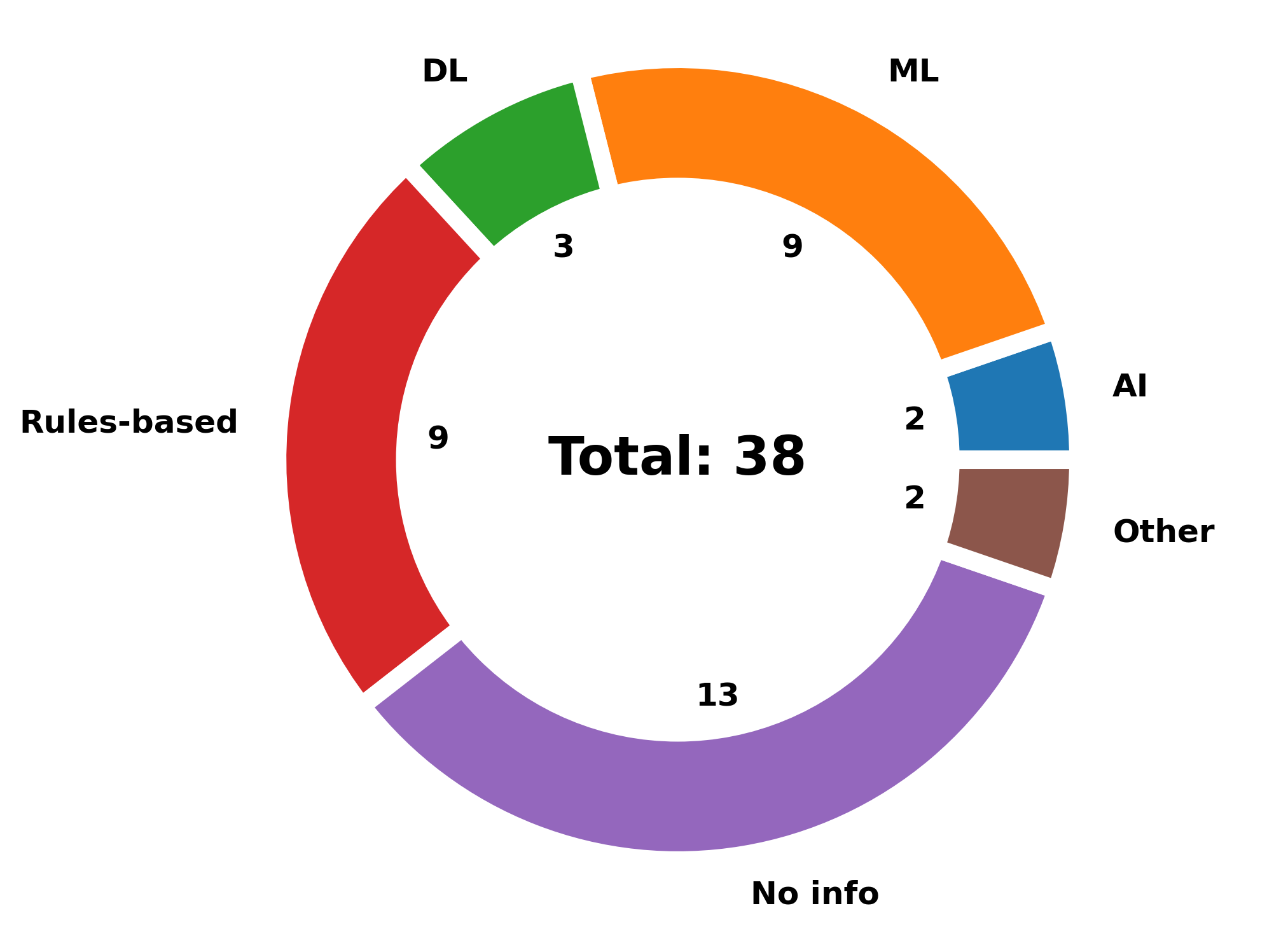}
    \caption{Categorization distribution of reviewed cybersecurity solutions.}
    \label{fig:solutionspie}
\end{figure}

Out of the 38 frameworks, we categorized two as \emph{AI}, three as \emph{DL}, nine as \emph{ML}, and nine as \emph{Rules-based}, while 13 provided not enough information. Two were discarded after further review. \Cref{fig:solutionspie} shows a graphical representation of the categorization results, while further details can be found in the appendix in \Cref{tab:frameworks}. The three products classified as Deep Learning are examined in more detail in \Cref{ssec:frameworkdescriptions} to provide an answer to RQ3. 

\subsection{Cybersecurity Frameworks Using Deep Learning}
\label{ssec:frameworkdescriptions}

To assess the state of deep learning in cybersecurity frameworks, we examine the three frameworks we identified for using deep learning: NVIDIA Morpheus, Vectra AI, and Check Point's R81. Following a brief introduction in this section, we look at the techniques these frameworks employ in \Cref{ssec:emplTechniques}.

\paragraph{NVIDIA Morpheus.}
NVIDIA Morpheus~\cite{nvidia_nvidia_2023} is an open-source cybersecurity framework that utilizes various machine-learning and deep-learning methods to enhance cybersecurity. It enables developers to create optimized AI pipelines for real-time data processing, classification, and threat detection. Morpheus integrates various tools, including pre-trained AI capabilities and telemetry analysis, to facilitate quick development and deployment of cybersecurity solutions. They claim, that with real-time telemetry and GPU acceleration, Morpheus can efficiently capture and act upon threats that were previously undetectable, providing valuable insights for immediate response. Morpheus makes use of the RAPIDS libraries for preprocessing, as discussed in \Cref{ssec:GPUsupport}.

Morpheus lists several use cases for its AI approach: Digital Fingerprinting, Phishing Detection, Sensitive Information Detection, Crypto-Mining Malware Detection, Ransomware Detection, and Fraudulent Transaction and Identity Detection. For this, Morpheus offers users the possibility of deploying their own models using NVIDIA's Triton Inference Server\footnote{\label{fn:triton}\url{https://developer.nvidia.com/triton-inference-server}}, which supports a variety of deep-learning frameworks, such as TensorFlow\cref{fn:tensorflow} or PyTorch\cref{fn:pytorch}~\cite{nvidia_nvidia_2023-1}. Currently, NVIDIA provides five pre-trained models each responsible for one cybersecurity task. \Cref{tab:morpheusmodels} shows the list of tasks and short descriptions, including used frameworks like XGBoost\footnote{\label{fn:xgboost}\url{https://xgboost.readthedocs.io/en/stable/}} and GraphSAGE\footnote{\label{fn:graphsage}\url{https://github.com/williamleif/GraphSAGE}}. 

\begin{table}[ht]
\centering
    \caption{NVIDIA Morpheus pre-trained models utilized for various cybersecurity tasks~\cite{nvidia_morpheus_2023}.}
    \label{tab:morpheusmodels}
    \begin{tabular}{@{}ll@{}}
        \toprule
        \textbf{Task}                         & \textbf{Description}                                                                                                                        \\ \midrule
        Anomalous Behavior Profiling & \multicolumn{1}{p{0.6\linewidth}}{\raggedright XGBoost model that classifies GPU behavior as normal or anomalous (e.g., crypto mining).} \\[0.6cm]
        Digital Fingerprinting       & \multicolumn{1}{p{0.6\linewidth}}{\raggedright Ensemble learning model consisting of an Autoencoder and fast Fourier transform reconstruction to detect changes in user behavior.} \\[0.6cm]
        Fraud Detection              & \multicolumn{1}{p{0.6\linewidth}}{\raggedright GraphSAGE along with XGBoost for fraud detection in a credit card transaction graph. }          \\[0.6cm]
        Ransomware Detection         & \multicolumn{1}{p{0.6\linewidth}}{\raggedright Random Forest model that classifies processes as ransomware or benign using volatile memory data. }                                  \\[0.6cm]
        Flexible Log Parsing         & \multicolumn{1}{p{0.6\linewidth}}{\raggedright Parsing of HTTP server logs with an undisclosed model. }                                                                             \\ \bottomrule
    \end{tabular}
\end{table}

While no explicit information about the datasets used to train, or federated learning capabilities is available in the documentation, Morpheus does stress the performance benefits stemming from the use of underlying NVIDIA technology. This includes aforementioned software, like Triton\cref{fn:triton}, but also NVIDIA hardware products NVIDIA BlueField Data Processing Units\footnote{\label{bluefield}\url{https://www.nvidia.com/en-us/networking/products/data-processing-unit/}}, for which their products are optimized. Furthermore, Morpheus provides a rich graphical user interface, that aims to give the user a quick overview of imminent threats, detected by, for example, their digital fingerprinting methods.

\paragraph{Vectra AI.}
Vectra~\cite{vectra_ai_2022} is a cybersecurity platform that leverages AI for advanced threat detection and response. Its approach focuses on identifying attacker methods, employing tactics, techniques, and procedures rather than isolated exploits. By optimizing AI models and leveraging streaming data analysis, Vectra enables real-time threat detection, allowing security teams to respond effectively. The platform's correlation algorithms analyze behaviors across various domains, attributing them to stable anchors like accounts or host machines. This so-called security-led approach enhances incident analysis, aiding in the identification and prioritization of progressing attacks. Vectra claims its methodology provides businesses with resilient and comprehensive protection against diverse cyber threats.

Vectra has presented three distinct applications of AI in their software through their whitepaper. These include threat correlation, detection of \ac{C2} channels, and identifying instances of privilege credential abuse. Threats are correlated from various artifacts collected from network metadata to identify attacks, while \ac{C2} channels are detected by examining the shape of network traffic. To detect privilege credential abuse, they build up user profiles and compare historic interactions with current user interactions flagging strong discrepancies as potential security violations.
They also equip their software with a dedicated UI that shows details on the various cybersecurity tasks. 

\paragraph{Check Point Quantum Cyber Security Platform R81.}
The Quantum Cyber Security Platform R81~\cite{check_point_r81_2022} is a cybersecurity framework that claims to be ``the industry's most advanced Threat Prevention and security management software''~\cite{check_point_r81_2022}. Its autonomous threat prevention system, Infinity Threat Prevention, seeks to reduce administrative overhead while enhancing an organization's security posture. By providing tailored policy profiles and enabling centralized security administration, R81 aims to ensure a unified view of security across networks, endpoints, mobile devices, IoT devices, and various cloud environments. Their release article states that the platform contains ``AI Deep Learning [that] prevents 5x more DNS attacks in real-time''~\cite{check_point_check_2023}. However, a more detailed description of these deep-learning techniques is not found.

\subsection{Employed Deep-Learning Techniques}
\label{ssec:emplTechniques}

After carefully reviewing the documentation and product description, we have identified a total of \emph{two} deep-learning algorithms that are utilized by the three cybersecurity frameworks. For instance, for NVIDIA Morpheus' digital fingerprinting task, an \emph{Autoencoder} algorithm is employed. Similarly, Vectra makes use of a \emph{\ac{LSTM}} model to efficiently detect \ac{C2} channels. 

All other techniques used by the frameworks are either not further disclosed, e.g., deep-learning approaches by the Checkpoint Quantum Cyber Security Platform R81, or are considered \ac{ML} techniques, e.g., Vectra's use of \ac{HDBSCAN}, which is a hierarchical clustering algorithm~\cite{campello_density-based_2013}.

\section{Conclusion \& Outlook}
\label{sec:conclusion}

In conclusion, our research on cybersecurity software frameworks reveals several key findings and insights. Firstly, we have observed that only a limited subset of cybersecurity solutions provides public information or technical details, which hinders transparency and collaboration in the field. This lack of openness restricts the broader community from fully understanding and evaluating the effectiveness of these solutions. Furthermore, our investigation indicates that \ac{DL} is not widely adopted in existing cybersecurity frameworks, and neither is GPU support. While there are promising ideas and research focusing on the utilization of \ac{DL} techniques such as \acp{CNN}~\cite{le_imids_2022,pinhero_malware_2021}, and Deep Neural Networks~\cite{xu_deep_2021}, their practical implementation and real-world viability remain uncertain. Deep-learning methods, although exhibiting potential, have not yet been proven ready for deployment in real-life environments due to underlying challenges, such as the availability of quality cybersecurity datasets or training and inference speed. Extensive testing and evaluation are necessary to determine if methods presented in research~\cite{berman_survey_2019} have efficacy and advantages over traditional methods. 

Looking ahead, future work in \ac{DL} methods for use in cybersecurity frameworks should consider focusing on addressing foundational problems within specific areas of cybersecurity, such as Intrusion Detection. For example, by providing better datasets for the training of \ac{ML} algorithms, it becomes possible to compare new approaches and train deeper models effectively. While \ac{DL} holds promise for enhancing cybersecurity, further research is needed to refine and adapt these techniques to address the unique challenges of the field. While overcoming issues such as interpretability, model explainability, and adversarial attacks is crucial to demonstrating the readiness of deep learning for deployment in real-life environments, researchers should also consider evaluating the operational viability of their approach. This includes metrics such as inference speed and hardware requirements, among others, and ensures that more theoretical solutions can be used in practice.

\section{Acknowledgement}
\label{sec:Acknowledgement}


\textit{<blinded>}

\printbibliography


\appendix

\newpage
\section{Overview of Reviewed Frameworks}

\begin{table}[ht]
    \centering
    \begin{tabular}{lll}
        \toprule
            \textbf{Framework} & \textbf{Category} & \textbf{Source} \\ 
        \midrule 
        Claroty & Rules-based & \cite{claroty_claroty_2022} \\ 
        Crowdstrike Network Detection & Rules-based & \cite{crowdstrike_network_2023} \\ 
        Datadog & Rules-based & \cite{datadog_detection_nodate} \\ 
        IBM Cloud Pak for Security QRadar Suite Threat Management & Rules-based & \cite{ibm_qradar_2023} \\ 
        Microsoft Sentinel & Rules-based & \cite{microsoft_what_2023} \\ 
        Rapid7 InsightIDR Feature Embedded Threat Intelligence & Rules-based & \cite{rapid7_insightidr_nodate} \\ 
        Snort & Rules-based & \cite{snort_team_snort_nodate} \\ 
        Suricata & Rules-based & \cite{suricata_suricata_2023} \\ 
        Zeek & Rules-based & \cite{zeek_zeek_2023} \\ 
        ExtraHop Reveal(x) & ML & \cite{extrahop_extrahop_2021} \\ 
        OSSEC OSSEC+ & ML & \cite{ossec_get_nodate} \\ 
        Progress WhatsUp Gold & ML & \cite{progress_using_2022} \\ 
        Palo Alto Networks Cortex XDR & ML & \cite{palo_alto_networks_cortex_2022} \\ 
        Skyhigh Security Cloud Access Security Broker (CASB) & ML & \cite{skyhigh_security_skyhigh_2022} \\ 
        Symantec Content Analysis & ML & \cite{symantec_symantec_2020} \\ 
        Trellix Network Detection and Response & ML & \cite{trellix_trellix_2023} \\ 
        VMware Contexa & ML & \cite{vmware_vmware_2022} \\ 
        Zscaler & ML & \cite{zscaler_solution_2021,zscaler_zscaler_2023} \\ 
        Check Point Quantum Cyber Security Platform: R81 & DL & \cite{check_point_check_2023,check_point_r81_2022} \\ 
        NVIDIA Morpheus & DL & \cite{allen_fingerprinting_2021,nvidia_nvidia_2023,nvidia_morpheus_2023,nvidia_nvidia_2023-1} \\ 
        Vectra AI & DL & \cite{vectra_ai_2022,vectra_fit_2021} \\ 
        Darktrace & AI & \cite{darktrace_reducing_2023} \\ 
        SentinelOne Complete \& Ranger & AI & \cite{sentinelone_sentinelone_nodate} \\ 
        GCP Chronicle & Other & \cite{ghanizada_autonomic_nodate} \\ 
        Oracle CASB Cloud Service & Other & \cite{oracle_oracle_nodate} \\ 
        Cisco SecureX & No info & \cite{cisco_cisco_2021} \\ 
        Cyberark & No info & \cite{cyberark_identity_nodate} \\ 
        Defendify & No info & \cite{defendify_cybersecurity_nodate} \\ 
        Fortinet FortiGate & No info & \cite{fortinet_fortigate_2023} \\ 
        Fotra Tripwire Enterprise & No info & \cite{tripwire_tripwire_nodate} \\ 
        Fujitsu Enterprise Cyber Security Solutions & No info & \cite{fujitsu_intelligence-led_nodate} \\ 
        FireEye Endpoint Threat Prevention HX Series (FireEye Security Platform) & No info & \cite{fireeye_hx_2014} \\ 
        Lookout & No info & \cite{lookout_lookout_2023,lookout_lookout_2023-1} \\ 
        McAfee Antivirus & No info & \cite{mcaffee_live_nodate} \\ 
        OpenText EnCase Endpoint Security & No info & \cite{opentext_opentext_2022} \\ 
        RSA ECAT & No info & \cite{rsa_rsa_2017} \\ 
        Trend Micro Service One & No info & \cite{trend_micro_trend_2023} \\ 
        VMware Carbon Black & No info & \cite{vmware_datasheet_2020} \\ 
        \bottomrule
    \end{tabular}
    \caption{Reviewed cybersecurity frameworks.}
    \label{tab:frameworks}
\end{table}

\end{document}